# Discovering the nanoscale origins of localized corrosion in additive manufactured stainless steel 316L by in situ liquid cell TEM

Mengkun Tian[1*], Jahnavi Desai Choundraj[2], Thomas Voisin[3], Y. Morris Wang[4], and Josh Kacher[2*]

1. Institute for Electronics and Nanotechnology, Georgia Institute of Technology, Atlanta, GA, USA
2. School of Materials Science and Engineering, Georgia Institute of Technology, Atlanta, GA, USA
3. Materials Science Division, Lawrence Livermore National Laboratory, Livermore, CA, USA
4. Department of Materials Science and Engineering, University of California Los Angeles, CA, USA

**Abstract**

Additive manufacturing (AM) by laser powder bed fusion (L-PBF) leads to the formation of a rich, hierarchical microstructure, including dislocation cell structures and elemental segregation. This structure has profound impacts on the corrosion behavior and mechanical properties of printed materials. In this study, we use *in situ* liquid cell scanning transmission electron microscope (STEM) to directly characterize the nanoscale origins of corrosion initiation in AM 316L stainless steel. Under applied anodic potentials, we found that the dislocation cellular boundaries were preferentially corroded and that pit-like features formed along the cellular boundaries. We directly observed the earliest stages of corrosion by controlling the biasing parameters to decelerate the corrosion processes. The results show that highly localized corrosion occurs via inclusion dissolution along dislocation cell boundaries. More widespread corrosion initiates at the dislocation cell boundaries and spreads throughout the dislocation networks.

## 1. Introduction

Research into and applications of additive manufacturing (AM) for metals and alloys has grown at an increasingly rapid pace in recent years. Along with the geometric flexibility inherent to AM processes, metals and alloys processed by laser powder bed fusion (L-PBF) AM have also been found to have exceptional mechanical properties and unusual corrosion behavior. [1-5] These properties have been ascribed to the rich hierarchy of dislocation and grain boundary structures formed during the manufacturing process. One of the most pronounced of the AM-induced microstructural features is a dense cellular dislocation network formed by repeated thermal loading cycles during printing, with cell spacings on the order of 100's of nanometers.[6],[7] Depending on processing conditions, these networks can be accompanied by elemental segregation, typically Cr and Mo for stainless steel, as well as the formation of nanoscale inclusions.

Corrosion studies of L-PBF AM stainless steel are still in the early stage[8] and conflicting reports on the susceptibility to and mechanisms of corrosion can be found in the literature. Two primary factors contribute to these disagreements: 1) the high dependency of the microstructure of AM parts on printing conditions leading to large performance variations in nominally identical printed materials and 2) the complexity of the microstructure induced by the printing process. Multiple reports have shown that a smaller crystal domain sizes, leading to a higher density of grain boundaries, leads to the formation of a more stable and thicker passive film in AM stainless steel compared to wrought materials. [9-11] In addition, it is believed[12] that the high dislocation-density and large lattice distortions at cellular boundaries further promote the formation of a stable passive film. However, Wang et al.[13] found that highly deformed nano-crystalline stainless steel with high dislocation density has worse corrosion behavior when exposed to a 3.5% NaCl solution than its annealed counterpart. They reported[13] that the dislocations weakened the bonding between the passive film and the substrate. In addition to the dense dislocation structures and refined grain size in AM steels, the elemental distribution in L-PBF processed alloys differs from wrought materials, including Cr and Mo segregation to the dislocation structures and the absence of MnS inclusions. The Cr and Mo segregation has been postulated to enhance the corrosion resistance of AM stainless steel as micro-galvanic coupling leads to a thicker and more stable passive layer[6]. Multiple groups have investigated the corrosion behavior of AM 316L in $Cl^-$ rich solution such as NaCl solution[14] and reagents with high concentration HCl such as Vilella's reagent[15, 16] and Kallinge's reagent[17]. Kong et al.[11] found using scanning kelvin probe force microscopy (SKPFM) that the cellular boundaries were approximately 5mV more positive than cell interiors and attributed this variation to the Cr and Mo segregation to the cell boundaries.

They also found that oxide and sulfide particles decreased the corrosion resistance in the AM steel parts. Nie et al.[18] found the pitting resistance of AM 316L was worse than the wrought 316L, attributing this observation to low OH$^-$ concentration at the boundaries coupled with micro-galvanic corrosion, which contributed to rapid pit formation. Similarly, other groups showed that the segregation of Mo[16] or Cr[17] in AM 316L would decrease the pitting resistance. In contrast, Chao *et al.* [19] found that AM austenitic stainless steel had improved pitting resistance, attributing this to the lack of MnS inclusion. This improved corrosion resistance was reversed upon annealing the material post-printing as this led to the formation of MnS inclusions.[20] However, correlations between microstructure, elemental segregation, and corrosion resistance in these studies relied on *post mortem* observations could not directly resolve the corrosion initiation processes.

*In situ* liquid cell transmission electron microscopy (LC-TEM) provides an avenue for directly resolving corrosion of pit formation processes in real time and at the nanoscale[21],[22-25]. As the measurements are conducted in the electron microscope, the corrosion observations can be directly related to defect structures, inclusions, and elemental segregation. In this paper, we apply this technique to investigate corrosion initiation in AM 316L in 1 w.t.% HCl. The AM 316L material investigated was fabricated by a Concept M2 using machine (Methods), via a cube geometry and an island scan strategy.[5, 26]

## 2. Materials and methods

2.1 Raw materials

AM 316L stainless steel cubes [10cm × 10cm × 8cm (height)] used in this work were produced with a Concept M2 (54μm beam size) laser powder bed fusion machine. The powder source was CL20 ES (Concept, Germany). The key laser parameters include laser power of 350 W, scan speeds 1700-1900 mm/s, a hatch spacing of 105 μm, and a build layer thickness of 30 μm. An island scan strategy was used. The as-built samples all have density >99.4%. The corrosion results reported in this study was obtained from the build-plane microstructures.

2.2 TEM sample preparation

The TEM sample was fabricated using a FEI Nova Nanolab 200 dual beam FIB/SEM. The TEM lamella was thinned down to 100 nm-200 nm thickness by FIB, and then transferred onto the working electrode of the chip. To avoid ion beam damage on the SiN$_x$ membrane, e-beam deposition was used for Pt deposition and welding the lamella onto the SiN$_x$ membrane. After the lamella was firmly attached to the membrane, a small trench (100nm in width) was cut on the lamella to separate it from the manipulator. To avoid possible leakage from this trench, Pt was deposited via e-beam deposition to seal the region with a controlled height slightly higher than the electrode. The manipulator was maintained in contact with the chip to reduce the charging effect during e-beam deposition

2.3 TEM imaging and EDS mapping

*In situ* TEM tests were performed in an FEI Tecnai F30 transmission electron microscope operated at 300kV using a Protochip Poseidon select holder. This TEM is equipped with an Oxford X-ray energy dispersive spectroscopy (EDS) detector.

*Ex situ* EDS mapping was conducted in a Hitach HD2700 aberration corrected STEM operated at 200 kV with a Bruker detector with ~0.3 steradian solid angle.

2.4 Electrochemical test

The samples used for electrochemical experiments are epoxy mounted with screws embedded at the rear of the samples for electrical connections. Prior to the experiments, the sample was polished using 800-4000 grit paper followed by diamond suspensions from 9μm to 1μm and a final polish using colloidal silica. Cyclic polarization scans were conducted using a three-electrode system with Pt as the counter electrode, saturated calomel electrode as the reference electrode, and the AM steel sample as the working electrode. Polarization scans were performed at a scan rate of 10 mV/s with the scan ranging from -0.2 V to +1.2 V with respect to the open circuit potential (OCP). The electrolyte used was 1 wt.% HCl solution, identical to the one used for the *in situ* experiments. Tescan Mira3 XM scanning electron microscope (SEM) was used to observe the microstructure of the sample post polarization scans to characterize the corrosion attack.

## 3. Experimental results and discussion
### 3.1. Characterization before LC test

Figs. 1a-b shows the liquid cell chips used for mounting and testing the samples. In this setup, the sample is isolated from the surrounding vacuum by two Si chips with 50 nm thick SiNx windows separated by 500 nm spacers. The top chip has embedded Pt electrodes, to which a focused ion beam (FIB)-machined sample is directly attached using a Pt weld. Microfluidics channels allow liquid flow over the chip at controllable rates while applying a bias to the sample itself. Prior to exposing the samples to a liquid environment, we conducted annular dark field (ADF) in the scanning transmission electron microscope (STEM) imaging and energy dispersive X-ray spectroscopy (EDS) to determine the initial microstructure and elemental distribution (Figs. 1c-e).

The STEM-ADF image shows that the sample is composed of dislocation cellular structures, with each cell having an average diameter of approximately 700 nm. There are also multiple inclusions in the sample, ranging from 20-100 nm, which appear dark in the STEM-ADF image. Although we were not able to resolve the elemental composition of these inclusions, prior analysis on this same material showed that they are transition-metal-rich silicates with varying levels of Mn.[5] EDS analysis shows that Ni, Cr, and Mo segregate to the dislocation cell boundaries, leading to depletion of these elements in the cell interiors. To highlight the small concentration difference of Ni, Cr, and Mo, we normalized the Fe peak in the EDS spectrum (Fig. 1e).

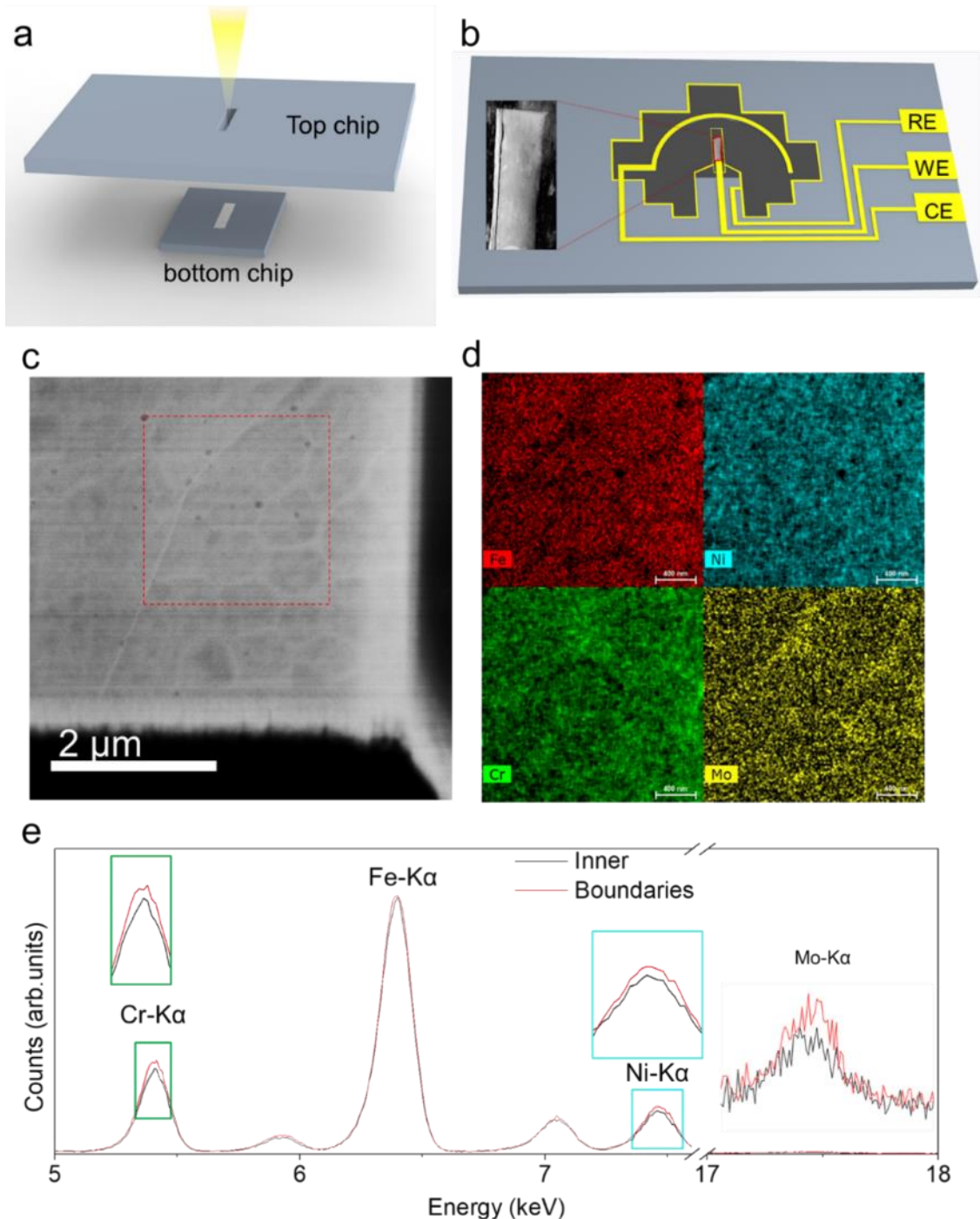

**Fig. 1 Schematic illustration of LC-TEM chips and STEM-EDS characterization of the dislocation cellular structure.** a) Schematic of the assembly of the chips. The yellow cone indicates the electron beam travel path. b) Schematic of the bottom chip with embedded working electrode (WE), reference electrode (RE) and counter electrode (CE). The inset is taken from the center part of the $SiN_x$ window, showing the location of the TEM lamella. c) STEM-ADF image of a part of TEM lamella, showing dislocation cellular structure where the boundary is in brighter contrast. The red square represents the region where the EDS map was collected. d) Montage of EDS maps of Fe, Ni, Cr and Mo. e) EDS spectra taken from the boundaries and the interior of the cell. The intensities of Fe-Kα taken from boundaries and

interior have been normalized. The insets magnify the intensity difference between boundary and interior, showing Cr, Ni and Mo enrichment at the boundaries.

## 3.2 In-situ LC-TEM test

Before *in situ* LC-TEM testing, the chips were cleaned using a Hitachi Zone cleaner for 3 minutes to make them hydrophilic. The electrolyte used was a 1 wt.% HCl solution in deionized (DI) water. The electrochemical tests including OCP measurements and biasing were performed using a Gamry 600 potentiostat. Low dose imaging was used, with a dose rate of approximately 0.025 electron/(nm$^2$·s). To minimize the electron beam induced damage while applying potential, all STEM images were recorded after applying voltage and after rinsing the lamella using DI water for 10 minutes with the maximum allowed flow rate and the beam was blanked between image acquisitions. The STEM-ADF images were acquired over a ~10μmx10μm field of view with 0.7 μs integration time per pixel. No noticeable electron beam effects were observed even during 70 minutes continuous integration time. However, we noticed that if the integration time of each pixel is more than the order of 1 second in the EDS mapping acquisition, the sample will be damaged right after applying biasing. Even flushing the sample for 10 minutes, slight beam damage can be seen when performing EDS-mapping with pixel time >1s. Therefore, to minimize the beam effect on the sample, all the EDS-mapping were performed without any HCl solution.

Fig. 2 shows the initial results from the *in situ* LC-TEM corrosion experiments. STEM-ADF imaging before liquid was flowed over the chips clearly shows the presence of dense, dislocation bands, similar to the dislocation cell structures shown in Fig. 1c (Figs. 2a-b). After imaging, we flowed deionized (DI) water over the chip for 10 minutes to thoroughly wet the sample. We then switched the flow to a 1 wt.% HCl solution and measured the OCP, finding it to be -560±30mV. Figs. 2c-d shows the same location after applying a -100 mV bias (~460mV over-potential) for approximately 30 seconds. During OCP measurements and biasing, we blanked the electron beam to avoid any beam effects. As can be seen, the right side of the sample which was the thinnest part was almost etched away. Figs. 2b and 2d show a comparison of the same region before and after applying biasing. The bright contrast of the cellular boundaries shown in Fig. 2b changed to dark contrast in Fig. 2d, indicating the cellular boundaries were attacked preferentially. In contrast, cellular boundaries under strong HCl acid etchants are reported to be more corrosion resistant.[11, 12, 15-17] To our knowledge, no previous research have directly observed such boundary corrosion of AM stainless steel in HCl solution in TEM. In addition to boundary corrosion, we were able to identify 180 pit-like features in the images. Approximately 95% of these features were concentrated along the dislocation cell boundaries, with the remaining 5% located in cell interiors.

Two representative pit-like features are indicated by a red circle (cell boundary pit) and a yellow circle (cell interior) in Fig. 2d. Recent research showed that the AM stainless steel could be susceptible to pitting.[27] However, whether those pit-like structures are real pits need time-resolved in-situ LC test as following.

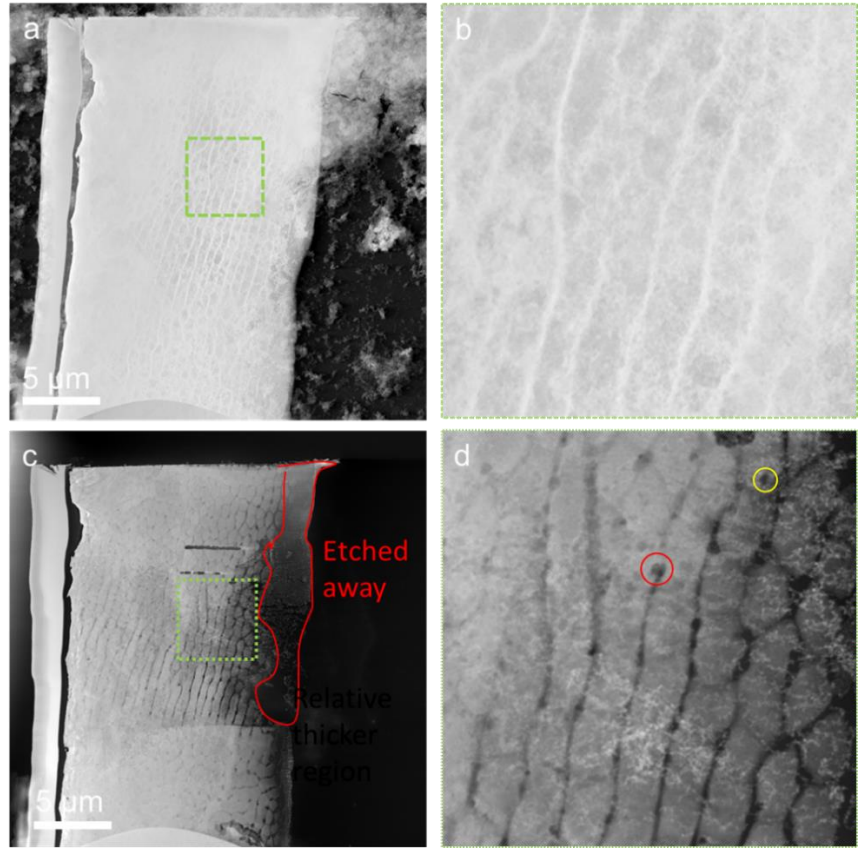

**Fig. 2 Comparison of the corrosion before/after biasing.** a) STEM-ADF image of the TEM lamella before biasing. b) Magnified image from the green dashed square in (a) showing the boundaries in brighter contrast. C) STEM-ADF image of the TEM lamella after biasing with etched away region indicated. D) Magnified image of the green dashed square in (d) showing the etched dislocation cell boundaries. Two regions of localized attack are highlighted by yellow and red circles showing pit-like features at a cell boundary (red) and cell interior (yellow).

### 3.3 Time resolved in-situ LC test

To decelerate the corrosion process and investigate the local corrosion mechanism, we adjusted the biasing potential and duration. Fig. 3a shows a STEM-ADF image of a lamella before exposure to the HCl solution. Two regions are highlighted for close inspection by red and green dashed squares. We selected these regions to track corrosion behavior at three different locations of interest: (1) at an inclusion at the intersection of dislocation cell boundaries and (2) at an inclusion located inside a dislocation cell interior. The center points of the two regions are highlighted by white dashed circles. We measured the OCP to be -454±7 mV, which is approximately 100 mV less than the OCP measured for the previous sample (-560±30 mV). The OCP differences between the two experiments (-454±7 mV vs -560±30 mV) originated from the flow rates (100µL/hr vs 300 µL/hr). [24] We then exposed the sample to HCl solution flow and applied a –400 mV potential, slightly higher than the OCP, for 4, 8, 12, 20 and 40 seconds. We blanked the electron beam again during applied bias to avoid any electron beam effects. As can be seen in Fig. 3b, taken after the 40 second interval, this over-potential did not initialize any corrosion. We then biased the sample at -300 mV for 4 seconds, resulting in the center of the red square to turn slightly darker, indicating localized attack at the inclusion site. The localized corrosion expanded somewhat over the course of 20 seconds at -300 mV (Fig. 3c-b). However, the corrosion front did not expand far beyond the initial inclusion area, suggesting that a dissolution process was occurring rather than pit formation. In Figs3.c-f, shortly after corrosion attack of the inclusion, the dislocation cell boundaries were also attacked and corroded, with the corrosion front rapidly spreading throughout the dislocation network. In comparison, the inclusion in the dislocation cell interior, highlighted by the green square in Fig. 3, was not attacked, suggesting that the inclusions are only susceptible to dissolution when embedded in the dislocation cell wall. This could be due either to the high strain energy associated with the dislocation accumulation or to the elemental segregation at the cell boundaries.

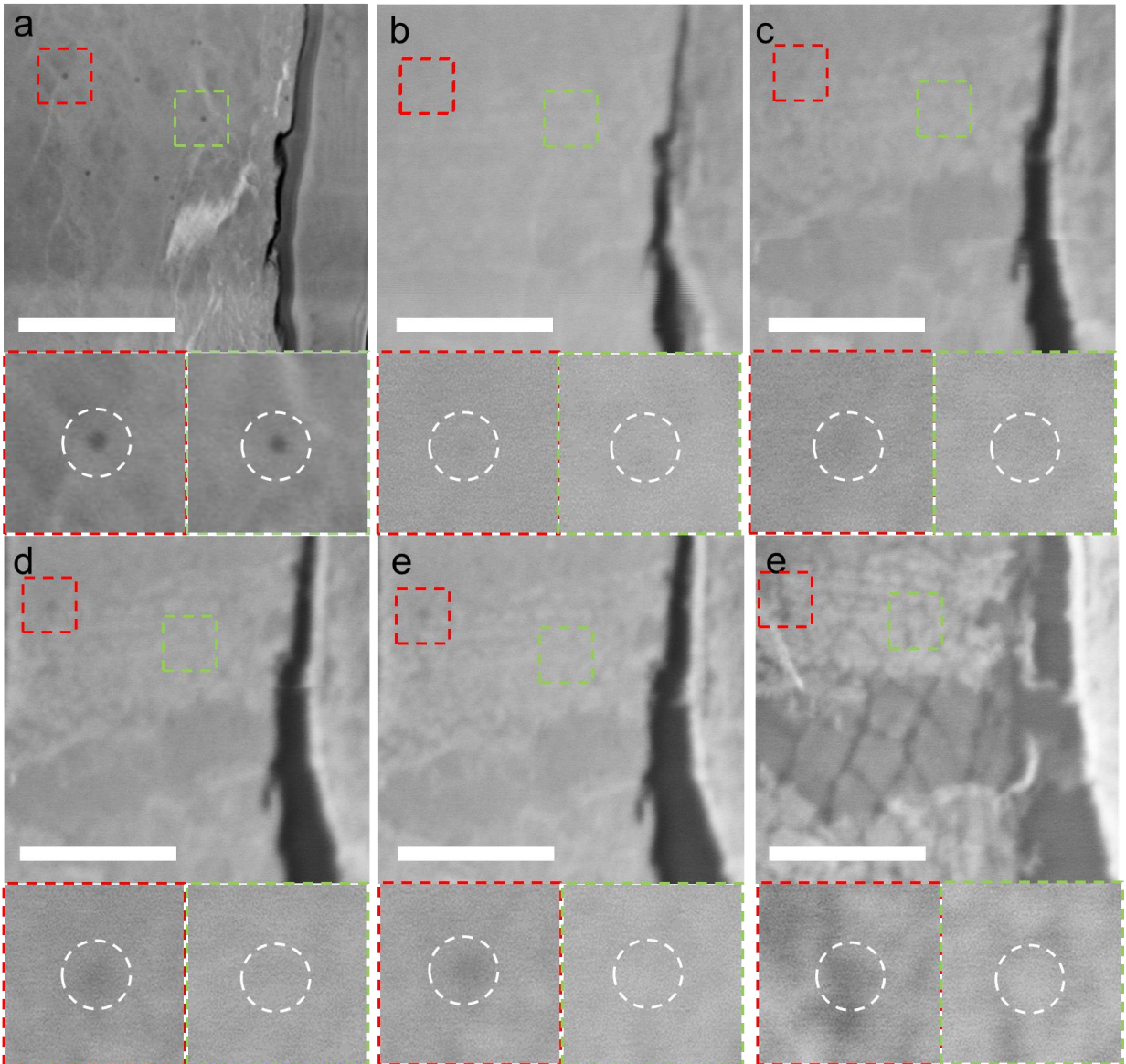

**Fig. 3 Monitoring the evolution of the morphology of the lamella during corrosion.** a) ADF-STEM image taken before sample was exposed to solution. b) Image taken after applying biasing at -400mV for 40s. c-f) Image taken after applying -300mV biasing for 4s, 9s, 14s and 20s, respectively. Highlighted show selective inclusion dissolution (red) and unattacked inclusion (green)Scale bar is 2µm.

### 3.4 Bulk electro-chemical test

To verify the observed behavior, bulk electro-chemical tests were run on the AM stainless steel sample. In agreement with the *in situ* TEM experiments, these tests showed that at high scan rates, 10mV/s, cellular attack was observed to dominate the response (Fig.4, scan data shown in Fig.S3). We also observed pit-like features distributed over the material surface, though primarily located along the dislocation cell networks, especially at the intersection points. By comparing the sizes of these pit-like features in the bulk SEM analysis and *in situ* TEM tests, we can see that they are approximately the same size, ranging from 70nm-210nm in the SEM images and 70nm-200nm in the TEM images. Considering high pitting potential (~900mV) from the polarization scan compared to the potential applied in the *in situ* experiments, this suggests that the pit-like features were not pits which should have significantly increasing sizes.. A systematic investigation of the influences that dictate cell boundary versus cell interior attack in bulk samples is currently ongoing. Here we take advantage of the unique liquid cell TEM capabilities to focus on the initial stages of attack.

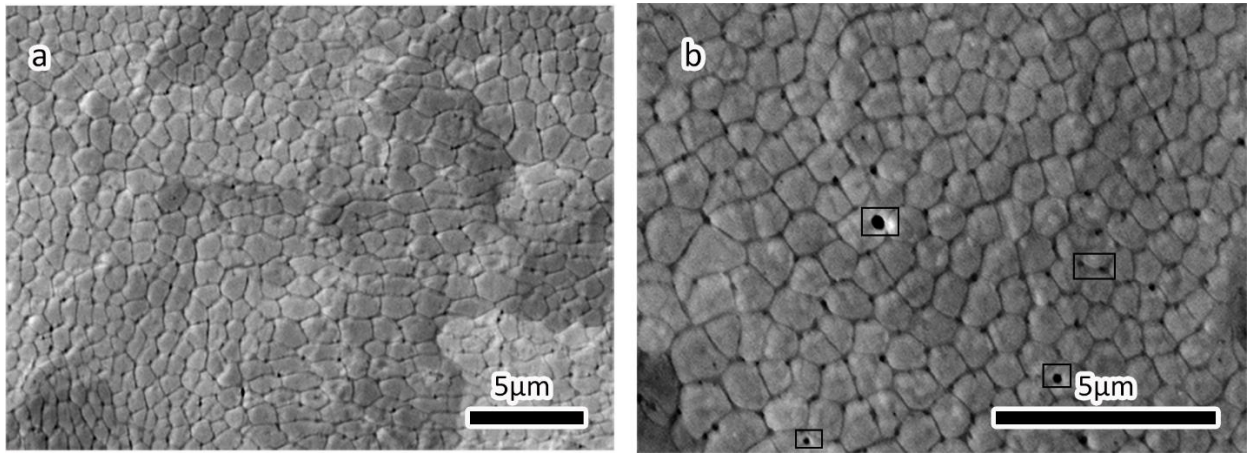

Fig.4 a and b SEM images of the AM Steel bulk sample after cyclic polarization scan showing cellular boundary attack and pits at the cellular boundaries and interiors.

## 4. Conclusion

In conclusion, we directly observed the earliest stages of corrosion attack in AM 316L stainless steel in 1wt.% HCl solution under biasing conditions using *in situ* liquid cell TEM. The results show that corrosion attack first takes place by dissolution of transition-metal-rich silicate inclusions, but does not spread beyond the inclusion sites and develop into pits which was further confirmed by the bulk corrosion test. The attacked inclusions were mostly located along dislocation cell boundaries, suggesting that either the localized strain fields or the elemental segregation to the boundaries increased the susceptibility to attack. After dissolution of the inclusions, we observed more widespread corrosion attack along the dislocation cell networks. These dislocation cells networks are rich in Cr and Mo relative to the surrounding matrix, suggesting that microgalvanic effects play an important role in the initial stages of corrosive attack in AM stainless steel.

## Declaration of Competing interests

The authors declare no competing interests.

## Author contributions

M.T. conducted the *ex-situ* and *in-situ* STEM works and FIB lift-out and transfer. J.D.C. performed the bulk electro-chemical test. T.V and Y.M.W fabricated the sample. All the author participated in manuscript writing.

## Acknowledgement

This research is supported by U.S. Office of Naval Research under Grant No. N00014-17-1-2646. This work was completed using the supported of IEN and MCF (as part of the Southeastern Nanotechnology Infrastructure Corridor, SENIC) at Georgia Tech, which are supported by the NSF through the National Nanotechnology Coordinated Infrastructure (NNCI). The work at Lawrence Livermore National Laboratory (LLNL) was performed under the auspices of the US Department of Energy under contract No. DE-AC52-07NA27344. The authors would like to thank Drs. W. King and C. Kamath (LLNL) for their early contributions to this work.

## Corresponding authors

Mengkun Tian: mtian37@gatech.edu
Josh Kacher: jkacher3@gatech.edu